\documentclass[superscriptaddress,onecolumn,floats,11pt,floatfix,nobibnotes,notitlepage,showkeys,longbibliography]{revtex4-2}

\usepackage{xcolor}
\usepackage[misc]{ifsym}
\usepackage{amsmath,amssymb,graphicx}
\usepackage{blindtext}
\usepackage{epstopdf}

\begin{document}
\title{Dzyaloshinskii-Moriya interaction as a fast quantum information scrambler}

\author{Fatih Ozaydin}
\email{(\Letter) fatih@tiu.ac.jp}
\affiliation{Institute for International Strategy, Tokyo International University, 1-13-1 Matoba-kita, Kawagoe, Saitama, 350-1197, Japan}
\affiliation{CERN, CH-1211 Geneva 23, Switzerland}

\author{Azmi Ali Altintas}
\email{altintas.azmiali@gmail.com}
\affiliation{Department of Physics, Faculty of Science, Istanbul University, 34116, Vezneciler, Istanbul, Turkey}

\author{Can Yesilyurt}
\email{can\texttt{-{}-}yesilyurt@hotmail.com}
\affiliation{Department of Physics, Faculty of Science, Istanbul University, 34116, Vezneciler, Istanbul, Turkey}

\author{Cihan Bay\i nd\i r}
\email{cbayindir@itu.edu.tr}
\affiliation{\.{I}stanbul Technical University, Engineering Faculty, 34469 Maslak, \.{I}stanbul, Turkey}  
\affiliation{Bo\u{g}azi\c{c}i University, Engineering Faculty, 34342 Bebek, \.{I}stanbul, Turkey}
\affiliation{CERN, CH-1211 Geneva 23, Switzerland}

\date{\today}

\begin{abstract}
Black holes are conjectured to be the fastest information scramblers, and within holographic duality, the speed of quantum information scrambling of thermal states of quantum systems is at the heart of studies of chaos and black hole dynamics.
Here, considering the Ising interaction on the thermal state of spin chains with Dzyaloshinskii-Moriya (DM) interaction and measuring the out-of-time-order correlation functions, we study the effect of DM interaction on the speed of scrambling the quantum information.
On the contrary to its advantages in quantum information and metrology such as exciting entanglement and quantum Fisher information, we show that DM interaction speeds up the information scrambling. 
We also show that the increasing temperature slows down the scrambling process due to vanishing quantum correlations. 
\end{abstract}


\maketitle
\section{Introduction}
The anisotropic antisymmetric Dzyaloshinskii-Moriya (DM) interaction  has been attracting a particular attention in quantum information science since Zhang found that it excites the entanglement of spin chains, or protects it against increasing temperature~\cite{Zhang07PRA}, motivating several other works on various quantum systems~\cite{Jafari08PRB,Kargarian09PRA,Mehran14PRA,Song14PhysA,Ozaydin2020LP}. 
Sharma et al. studied entanglement sudden death and birth~\cite{Sharma13QIP} and entanglement dynamics~\cite{Sharma14QIP} of qubit-qutrit systems, also of Werner states~\cite{Sharma15QIP} and qubit-qutrit systems with x-component DM interaction~\cite{Sharma16CTP}. 
Because, unlike entanglement measures, quantum Fisher information (QFI) is not an entanglement monotone that it can be increased via local operations and classical communications~\cite{Erol14SRep}, we asked whether it can be excited via DM interaction too, and showed that it can be~\cite{Ozaydin15SRep,Ozaydin20OQEL}.
Bound entanglement (BE)~\cite{Horodecki1998PRLIsThere} and its activation are among the most interesting phenomena in quantum mechanics~\cite{Horodecki1999PRL,Ozaydin-BoundZeno}. 
Along this vein, Sharma et al. showed that DM interaction can be used as an agent to free BE~\cite{Sharma16QIP}.
Besides these \textit{positive} influences on the quantum dynamics, Vahedi et al. showed that DM interaction can drive a quantum system into chaos~\cite{Vahedi16Chaos}, motivating us for the present work due to the close relation between information scrambling and chaos in quantum systems.

Measuring the out-of-time-correlation (OTOC) functions has recently opened new insights in the \textit{black hole in lab} studies, in particularly in quantum chaos within the holographic duality.
Through reversing the sign of the Hamiltonian of a many body system, Swingle et al. showed that scrambling of quantum information can be probed~\cite{Swingle16PRA}.
Dag et al. studied information scrambling and OTOCs in cold atoms~\cite{Ceren19PRA}, and also showed that OTOC functions can detect the quantum phases~\cite{Ceren19PRL}.
Very recently, Sharma et al. have established mathematical connections between quantum information scrambling and quantum correlation quantifiers through OTOC functions~\cite{Sharma21QIP}.

In this work, we investigate the influence of DM interaction on the speed of information scrambling in quantum systems by calculating the OTOC functions.
DM interaction has a good reputation in quantum information and computation tasks. 

However, finding that it scrambles quantum information faster could destroy this reputation in systems which exhibit chaotic dynamics in certain conditions. 
On the other hand, it has a potential to contribute to the \textit{black hole in lab} studies. 
In either way, it opens new insights in our understanding the fundamentals of quantum mechanics and also in developing quantum technologies.

\section{Physical Model and Information Scrambling}
For the interaction Hamiltonian $H_I$, we consider the following Ising model, which is a simplified version of the power-law quantum Ising model considered by Swingle and Halpern~\cite{Swingle18PRA} for a rapid scrambling at early times
\begin{equation}\label{eq:HamIsing}
	H_{\text{I}}=-\sum^{n-1}_{r=1}J \sigma^z_r \sigma^z_{r+1} - \sum^{n}_{r} h_x \sigma^x_r - \sum^{n}_{r} h_z \sigma^z_r,
\end{equation}

\noindent where the interaction-energy scale is chosen to be $J=-1$, the transverse field and the position-dependent longitudinal field are chosen to be $h^x=1.05$, and $h_r^z=0.375 (-1)^r$, respectively. 

Although this simplified version does not lead to such a rapid scrambling due to the exponential growth of OTOCs, it is sufficient to observe the influence of DM interaction on the scrambling.

The OTOC operators are chosen to be Pauli-x operator on the first and the last qubit, i.e. $V=\sigma_1^x$ and $W=\sigma_n^x$.
The operator $W$ is subject to a time evolution described by Hamiltonian $H$ as $W_t = U(-t) W U(t)$ where $U(t) = e^{-iHt}$ with $\hbar=1$.
We measure the out-of-time-correlation function through fidelity $\mathcal{F}$ of two instances of the initial state $\rho_i$ as $F(t)=\sqrt {Re[\mathcal{F}(\rho_a,\rho_b)]}$ with	
$\rho_a = W_t V \rho_i V^{\dagger} W_t^{\dagger}$ and 
$\rho_b = V W_t \rho_i W_t^{\dagger} V^{\dagger}$. 
Hence, the deviation of $F(t)$ from unity, i.e. the discrepancy between $\rho_a$ and $\rho_b$ detects the scrambling of quantum information.

\begin{figure}[t!]
	\centerline{
		\includegraphics[width=0.7\columnwidth]{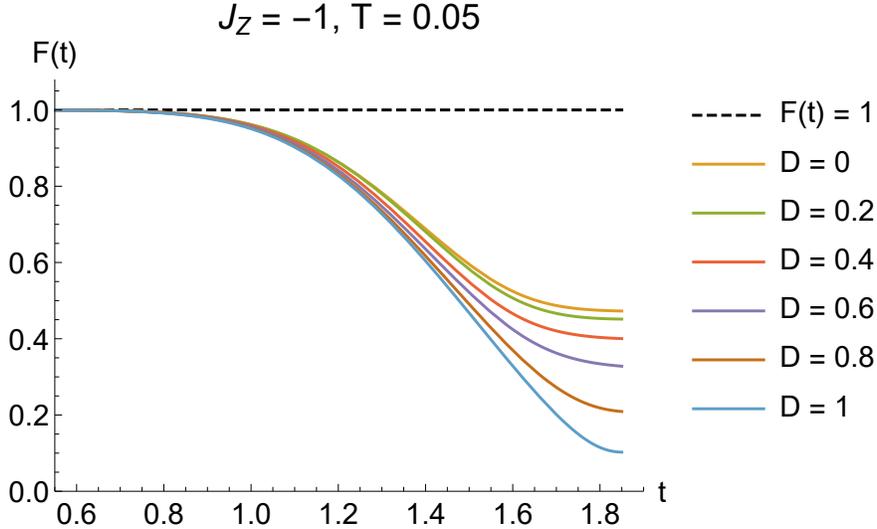} 
	}
	\caption{Information scrambling speed with respect to strength of Dzyaloshinskii-Moriya interaction.   }\label{fig:fig1}
\end{figure}

Along with such a spin-1/2 interaction model and OTOC operators, one can consider an arbitrary two-level quantum system to study information scrambling.
The system we consider in this work is a thermalized $n$ spin-1/2 chain with DM interaction, described by the Hamiltonian

\begin{equation}\label{eq:Ham}
	H_{\text{DM}}=\sum^{n-1}_{k=1} {1 \over 2} \left[ J_x \sigma^x_k \sigma^x_{k+1}
	+ J_y \sigma^y_k \sigma^y_{k+1}
	+ J_z \sigma^z_k \sigma^z_{k+1}
	+ \overrightarrow{D}  \cdot (\overrightarrow{\sigma}_{k} \times \overrightarrow{\sigma}_{k+1}) \right],
\end{equation}

\noindent where $J$ is the coupling constant, $D$ is the strength of DM interaction, and we choose $J_x = J_y$, $J_z < 0$ and $\overrightarrow{D}= D \overrightarrow{z}$.
The density matrix of the thermal entangled state of the spin-1/2 system which we use as the initial state $\rho_i$ is found as 
\begin{equation}
	\rho_i = e^{- H_{\text{DM}} / kT } / \text{Tr}(e^{- H_{\text{DM}} / kT }),	
\end{equation}

\noindent where $k$ is the Boltzman constant and $T$ is the temperature.

\begin{figure}[t!]
	\centerline{
		\includegraphics[width=0.7\columnwidth]{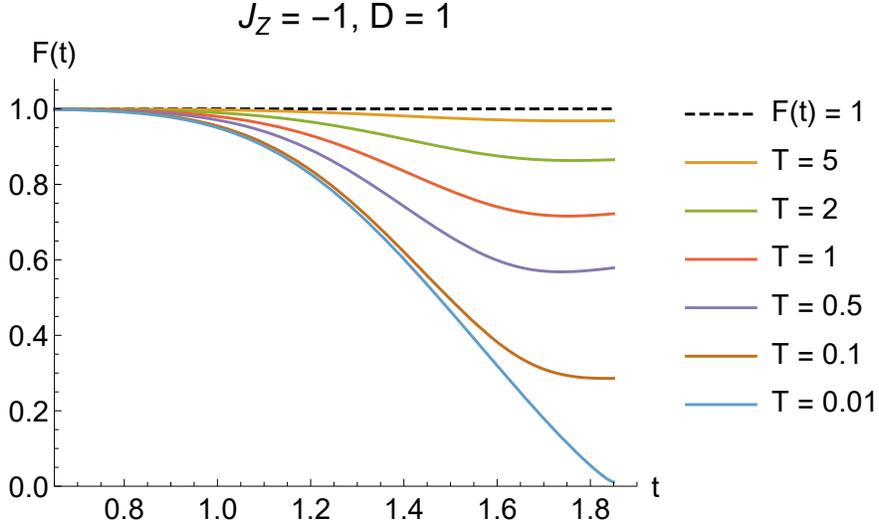} 
	}
	\caption{Information scrambling speed with respect to strength of Dzyaloshinskii-Moriya interaction.   }\label{fig:fig2}
\end{figure}

Since our major question in this work is whether DM interaction speeds up quantum information scrambling, we first set $J_z = -1$ and choose $T = 0.05$.
For a set of values of the strength of DM interaction ranging from $D = 0$ to $D = 1$, we plot $F(t)$ in Fig.~\ref{fig:fig1}, which clearly shows that as the strength of DM interaction increases, quantum information is scrambled faster.\\
\indent 
In the above result, in order to observe the effect of DM interaction solely, we kept the other system parameters constant. However, thermalization and decoherence effects are in the heart of scrambling in quantum systems.
Swan et al. show that fidelity OTOCs can shed light on the connections of scrambling, thermalization, entanglement and quantum chaos~\cite{SwanNatComm}.
In order to discriminate scrambling from decoherence, 
Landsman et al. implemented and verified scrambling on an ion trap quantum computer by designing a  teleportation circuit ~\cite{LandsmanNature}.
Because temperature is a key factor in the quantum dynamics of the thermal state of a spin chain, it is a natural direction to investigate how the increasing temperature affects the scrambling speed.
One might expect in general that the temperature speeds up the scrambling process. 
However, as can be seen from Fig.~\ref{fig:fig2} where we set $J_z = -1$ and $D = 1$, scrambling slows down with respect to the increasing temperature. 
The physical interpretation of this result is that as the temperature increases, as can be calculated easily, quantum correlations of the initial state decreases beyond the point that DM interaction can excite.
Hence, scrambling, which is a pure quantum mechanical effect tends to vanish. 
In other words, at high temperatures, there is no significant quantum information left to scramble.

\section{Conclusion}
In conclusion, we have studied the influence of DM interaction on the speed of information scrambling in quantum systems. 
Focusing on a spin-1/2 chain with DM interaction and considering Ising interaction, we have shown that DM interaction speeds up the quantum information scrambling.
Our work has a potential to contribute to  chaos and black hole dynamics within holographic duality.

\section*{Acknowledgments}
\noindent F.O. acknowledges the financial support of Tokyo International University Personal Research Fund. C.Y. acknowledges the Istanbul University Scientific Research Fund, Grant No. BAP-2019-33825.

\section*{Compliance with ethical standards}
\noindent \textbf{Conflict of interest} We have no competing interests.

\noindent \textbf{Ethics statement} This work did not involve any active collection of human data.

\noindent \textbf{Data accessibility statement} This work does not have any experimental data.

\noindent \textbf{Funding} We have no competing financial interests.


\end{document}